\documentclass[11pt,a4paper]{article}

\usepackage{jheppub_kim}
\usepackage{rotating}
\usepackage{graphicx,epsfig}
\usepackage{amsmath}
\usepackage {amssymb}
\usepackage{subfigure}

\usepackage{relsize}
\usepackage{graphicx}
\usepackage{epsfig}
\usepackage{bm}
\usepackage{amsfonts}
\usepackage[latin1]{inputenc}
\usepackage{amssymb}
\usepackage{color}
\usepackage{float}
\usepackage{amsmath}
\usepackage{dcolumn}
\usepackage{hyperref}
\usepackage{array}
\usepackage{lscape}

\usepackage{array,multirow}
\usepackage{soul}
\usepackage{subfigure}
\usepackage{dsfont}
\usepackage{hyperref}
\usepackage{txfonts}
\usepackage{newlfont}
\usepackage{times}

\providecommand{\U}[1]{\protect\rule{.1in}{.1in}}

\def\be{\begin{equation}}
\def\ee{\end{equation}}
\def\bea{\begin{eqnarray}}
\def\eea{\end{eqnarray}}
\def\eqi{\begin{equation}}
\def\eqf{\end{equation}}
\def\eqia{\begin{eqnarray}}
\def\eqfa{\end{eqnarray}}

\title{  Updated constraints on $f(T)$ models using direct and indirect
measurements of the Hubble
parameter}
\author[a]{Spyros Basilakos}
\author[b]{Savvas Nesseris}
\author[a]{Fotios K. Anagnostopoulos}
\author[c,d,e]{Emmanuel N. Saridakis}
\affiliation[a]{Academy of Athens, Research Center for Astronomy and
Applied Mathematics, Soranou Efesiou 4, 11527, Athens, Greece}

\affiliation[b]{Instituto de F\'isica Te\'orica UAM-CSIC, Universidad Auton\'oma de
Madrid,
Cantoblanco, 28049 Madrid, Spain}

 \affiliation[c]{Department of Physics, National Technical University of Athens, Zografou
Campus GR
157 73, Athens, Greece}
\affiliation[d]{Chongqing University of Posts \& Telecommunications, Chongqing,
400065,China}

\affiliation[e]{CASPER, Physics Department, Baylor University, Waco, TX 76798-7310, USA}

\emailAdd{svasil@academyofathens.gr}
\emailAdd{savvas.nesseris@csic.es}
\emailAdd{fotis-anagnostopoulos@hotmail.com}
\emailAdd{Emmanuel\_Saridakis@baylor.edu}

\abstract{ We  extract observational constraints
on $f(T)$ gravity, using the recently proposed
statistical method which is not affected by
the value of $H_0$ and thus it bypasses the
problem of the disagreement in its exact numerical
value between Planck and direct
measurements. We use direct measurements of the Hubble parameter
with the corresponding covariance matrix, and for
completeness we perform a joint analysis
using the latest data from Supernovae type Ia based on JLA sample,
quasi-stellar
objects, and Cosmic Microwave Background shift parameter from Planck.
We analyze a large family of
$f(T)$ models, and  we compare the fitting results
with $\Lambda$CDM cosmology using
the AIC statistical test.
Utilizing only the Hubble parameter data we find that
in the case of the power-law $f(T)$ model
a small but non-zero deviation from
$\Lambda$CDM cosmology is slightly favored at 1-$\sigma$, nevertheless the
corresponding AIC value shows a statistical equivalence with it.
Finally, the join analysis reveals that
all $f(T)$ models are very efficient and in very good
agreement with observations.
}
\begin{document}
\maketitle

\section{Introduction}

The early and late time phases of accelerated expansion are amongst the
most interesting findings of modern cosmology. In general, one may follow two main
directions in order to offer an explanation. The first way  is to keep general
relativity as the theory of gravitational interactions and introduce new components
such as the dark energy sector \cite{Copeland:2006wr,Cai:2009zp} and/or the inflation
field(s) \cite{Starobinsky:1980te,Inflation1,Inflation2}. The second direction is to
construct a modified
theory of
gravity, which possess general relativity as a particular limit, but with additional
degrees of freedom  that can drive acceleration
\cite{Copeland:2006wr,Sahni:2006pa,modgrav1,modgrav2,Capozziello:2011et}.

Most works in modified theories of gravity are based on the usual curvature-based
formulation, and  modify in various ways the Einstein-Hilbert action, as for instance
in $f(R)$ gravity \cite{DeFelice:2010aj,Nojiri:2010wj}. However, one
can equally well construct  gravitational modifications starting from the
torsional formulation of gravity, and in particular from the Teleparallel Equivalent of
General Relativity (TEGR)
\cite{ein28,Hayashi79,HayashiAddendum,Pereira.book,Maluf:2013gaa}. Since
in
this
framework the gravitational Lagrangian is the torsion scalar $T$, the simplest torsional
modified gravity would be to extend $T$ to $f(T)$, obtaining $f(T)$ gravity
\cite{Bengochea:2008gz,Linder:2010py} (see \cite{Cai:2015emx} for a review). Although
TEGR is completely equivalent with general relativity at the level of equations, $f(T)$
gravity is different from $f(R)$ gravity, namely it is a novel class of gravitational
modification, and that is why it has attracted the interest of the literature. The
significant theoretical advantage of $f(T)$ gravity is that its field equations are
always second-order, in contrast with $f(R)$ and other
curvature-based modified theories of
gravity. Concerning the cosmological implications, $f(T)$
gravity proves to lead to interesting phenomenology at both  early
\cite{Ferraro:2006jd,Ferraro:2008ey}, as well as at late times
\cite{Dent:2011zz,Zheng:2010am,Zhang:2011qp,Cai:2011tc,Sharif001,Li:2011rn,
Chattopadhyay:2011fp,Capozziello:2011hj,Daouda:2011rt,Geng:2011aj,Wu:2011kh,Wei:2011aa,
Guo:2015qbt,Atazadeh:2011aa,Ong:2013qja,Amoros:2013nxa,Miao:2011ki,Otalora:2013dsa,
Bamba:2013jqa,Paliathanasis:2014iva,Harko:2014aja,Geng:2014nfa,Hanafy:2014ica,
Darabi:2014dla,
Fazlpour:2016bxo,Malekjani:2016mtm,Farrugia:2016qqe,Sk:2017ucb,Bahamonde:2017wwk,
Hohmann:2017jao,Conroy:2017yln,Hohmann:2018rwf}.

Similarly to many models of modified gravity one can use observational data in order to
constrain the large variety of $f(T)$ models that can be theoretically constructed.
Hence, in the literature one can have works that use Solar System data
\cite{Iorio:2012cm}, or data from  Supernovae type Ia
data  (SNIa),  Cosmic Microwave Background (CMB) shift parameters,
Baryonic Acoustic Oscillations (BAO)  growth rate and Hubble data observations
\cite{Wu:2010mn,Bengochea001,Cardone:2012xq,Nesseris:2013jea,Basilakos:2016xob,Nun16,
Nunes:2016plz,Capozziello:2015rda,Nunes:2018xbm,Xu:2018npu}
in order to
extract constraints on the involved model parameters. At the cosmological level the
above observations probe the integral of
the Hubble parameter, and hence in the various analyses one must insert the current value
of the Hubble function $H_0$. Unfortunately, the exact value of  $H_0$ is a matter of
debate due to the known disagreement between Planck \cite{Planck16}
and local supernovae type Ia
(hereafter SNIa; Riess et al. \cite{Riess16}) measurements.

 In order to alleviate the aforementioned issue of the $H_0$ value, in
\cite{Anagnostopoulos:2017iao} the authors presented a new  statistical method which is
not affected by the value of $H_0$.
In particular, using the direct measurements of the Hubble expansion
the so called $H(z)$ data \cite{Farooq:2016zwm},
they marginalized analytically the Hubble constant in the likelihood
function using Bayesian statistics. The latter approach has the property of circumvent
the
$H_0$ value problem and reducing the parameter space without adding numerical complexity.
In the present work we desire to apply this $H_0$-independent method
in order to impose
constraints on various $f(T)$ models.
%In this way, the obtained constraints and
%contour plots will be more precise than those extracted previously.
Additionally, for
completeness we perform a joint analysis using data
from standard candles such as SNIa,
and quasi-stellar objects (QSOs), and also CMB shift parameters from
Planck \cite{Ywang16}.
We would like to stress that this is the first time that
the $H_0$-independent method of the $H(z)$ data,
the covariance matrix of the $H(z)$ measurements
\cite{Yu17,Yu17b}, the JLA and the QSOs data
are
%used
combined with the Planck CMB shift parameter
%other probes
toward constraining the $f(T)$ models.

The plan of the work is as follows:
In section \ref{fTmodel} we review
$f(T)$ gravity and cosmology, and in section
\ref{specificfTmodels} we provide the basic
specific $f(T)$ models that have appeared in
the literature. In section \ref{observdata} we present the
method and the observational
data sets that we use and %in section \ref{results}
we perform a detailed observational
analysis providing the corresponding best fit values and contour plots for the various
$f(T)$
models. In section V,
based on Monte Carlo simulations, we quantify the ability
of future direct measurements of the Hubble parameter to place
strong constraints on the $f(T)$ models. Finally, we discuss
our conclusions in section VI.

\section{$f(T)$ gravity and cosmology}
\label{fTmodel}

In this section we briefly review $f(T)$ gravity and we apply it in a cosmological
framework. In torsional formalism it proves convenient to use as   dynamical variables
the vierbeins fields ${\mathbf{e}_A(x^\mu)}$, which form an orthonormal
basis for the tangent space at each point  of the manifold $x^\mu$  (namely
$\mathbf{e}_A\cdot \mathbf{e}_B=\eta_{AB}$,
with $\eta_{AB}={\text {diag}} (1,-1,-1,-1)$.
In a coordinate basis they are expressed   as
$\mathbf{e}_A=e^\mu_A\partial_\mu $ and therefore  the metric  is acquired as
\begin{equation}  \label{metricc}
g_{\mu\nu}(x)=\eta_{AB}\, e^A_\mu (x)\, e^B_\nu (x),
\end{equation}
where Greek indices are used for the coordinate space-time while Latin indices for the
tangent one.

One can now introduce the curvature-less Weitzenb\"{o}ck connection
$\overset{\mathbf{w}}{\Gamma}^\lambda_{\nu\mu}\equiv e^\lambda_A\:
\partial_\mu
e^A_\nu$ \cite{Weitzenb23}, in terms of which he can define the torsion tensor as
\begin{equation}
\label{torsten}
{T}^\lambda_{\:\mu\nu}=\overset{\mathbf{w}}{\Gamma}^\lambda_{
\nu\mu}-%
\overset{\mathbf{w}}{\Gamma}^\lambda_{\mu\nu}
=e^\lambda_A\:(\partial_\mu
e^A_\nu-\partial_\nu e^A_\mu),
\end{equation}
which contains all the information of the gravitational field. Contracting the torsion
tensor we obtain the torsion scalar as
\begin{equation}
\label{torsiscal}
T\equiv\frac{1}{4}
T^{\rho \mu \nu}
T_{\rho \mu \nu}
+\frac{1}{2}T^{\rho \mu \nu }T_{\nu \mu\rho }
-T_{\rho \mu }^{\ \ \rho }T_{\
\ \ \nu }^{\nu \mu },
\end{equation}
which is then used as the Lagrangian of teleparallel gravity (similarly to the use of the
 Ricci scalar as the Lagrangian of general relativity). Through variation of the
teleparallel action in terms of the vierbeins, one obtains exactly the same equations
with general relativity, and that is why
the theory at hand was named teleparallel equivalent
of general relativity (TEGR).

Inspired by the $f(R)$ extensions of general
relativity, one can generalize $T$ to
a function $T+f(T)$.
The resulting $f(T)$ theory of gravity is characterized
by the action
\cite{Cai:2015emx}
\begin{eqnarray}
\label{action0}
S = \frac{1}{16\pi G}\int d^4x e \left[T+f(T)\right],
\end{eqnarray}
where $e = \text{{det}}(e_{\mu}^A) = \sqrt{-g}$ and $G$ is the gravitational
constant (the light speed is set to 1 for simplicity).

In order to study the cosmological applications of $f(T)$
gravity first we must add the matter
and radiation sectors, and therefore the
total action  takes the form
\begin{eqnarray}
\label{actionfull}
 I = \frac{1}{16\pi G}\int d^4x e
\left[T+f(T)+L_m+L_r\right].
\end{eqnarray}
Varying the above action with
respect to the vierbeins we extract the field equations as
\begin{equation}\label{equationsom}
e^{-1}\partial_{\mu}(ee_A^{\rho}S_{\rho}{}^{\mu\nu})[1+f_{T}]
 +
e_A^{\rho}S_{\rho}{}^{\mu\nu}\partial_{\mu}({T})f_{TT}
-[1+f_{T}]e_{A}^{\lambda}T^{\rho}{}_{\mu\lambda}S_{\rho}{}^{\nu\mu}+\frac{1}{4} e_ { A
} ^ {
\nu
}[T+f({T})]
= 4\pi Ge_{A}^{\rho}\overset {\mathbf{em}}T_{\rho}{}^{\nu},
\end{equation}
where $f_{T}=\partial f/\partial T$, $f_{TT}=\partial^{2} f/\partial T^{2}$,
and with $\overset{\mathbf{em}}{T}_{\rho}{}^{\nu}$  denoting the matter energy-momentum
tensor. In the above equations we have introduced the ``super-potential''
$
S_\rho^{\:\:\:\mu\nu}\equiv\frac{1}{2}\Big(K^{\mu\nu}_{\:\:\:\:\rho}
+\delta^\mu_\rho
\:T^{\alpha\nu}_{\:\:\:\:\alpha}-\delta^\nu_\rho\:
T^{\alpha\mu}_{\:\:\:\:\alpha}\Big)$, with
$K^{\mu\nu}_{\:\:\:\:\rho}\equiv-\frac{1}{2}\Big(T^{\mu\nu}_{
\:\:\:\:\rho}
-T^{\nu\mu}_{\:\:\:\:\rho}-T_{\rho}^{\:\:\:\:\mu\nu}\Big)$ the con-torsion tensor.

As a second step we impose the homogeneous and isotropic geometry
$
e_{\mu}^A={\text
{diag}}(1,a,a,a)$,
which corresponds to the spatially
flat Friedmann-Robertson-Walker (FRW)  metric
\begin{equation}
ds^2= dt^2-a^2(t)\,  \delta_{ij} dx^i dx^j,
\end{equation}
where $a(t)$ is the scale factor. Inserting this vierbein choice
into the field equations
(\ref{equationsom}) we extract the Friedmann equations, namely
\begin{eqnarray}\label{Fr11}
&&H^2= \frac{8\pi G}{3}(\rho_m+\rho_r)
-\frac{f}{6}+\frac{Tf_T}{3}\\\label{Fr22}
&&\dot{H}=-\frac{4\pi G(\rho_m+P_m+\rho_r+P_r)}{1+f_{T}+2Tf_{TT}},
\end{eqnarray}
with
$H\equiv\dot{a}/a$   the Hubble function, and where dots denote
derivatives with respect to $t$ (note that we have used the fact that    $T=-6H^2$, which
 straightforwardly arises from (\ref{torsiscal}) in FRW geometry). Moreover, in the
above equations $\rho_m$, $\rho_r$ and   $P_m$, $P_r$ are
respectively the energy densities and pressures of the matter and radiation sectors,
considered to correspond to perfect fluids.

From the form of the Friedmann equations
(\ref{Fr11}) and (\ref{Fr22})
it is implied that we can define the energy density and pressure of the
effective dark energy
sector as
\begin{eqnarray}
&&\rho_{DE}\equiv\frac{3}{8\pi
G}\left[-\frac{f}{6}+\frac{Tf_T}{3}\right], \label{rhoDDE}\\
\label{pDE}
&&P_{DE}\equiv\frac{1}{16\pi G}\left[\frac{f-f_{T} T
+2T^2f_{TT}}{1+f_{T}+2Tf_{TT}}\right],
\end{eqnarray}
and furthermore we can write its effective equation-of-state parameter as
\begin{eqnarray}
\label{wefftotf}
 w\equiv\frac{P_{DE}}{\rho_{DE}}
=-\frac{f/T-f_{T}+2Tf_{TT}}{\left[1+f_{T}+2Tf_{TT}\right]\left[f/T-2f_{T}
\right] }.
\end{eqnarray}
Lastly, the equations close by   considering the conservation  equations  of the matter
and radiation sectors, namely
\begin{eqnarray}
\label{mattradevol}
 \dot{\rho}_m+3H(\rho_m+P_m)=0\\
  \dot{\rho}_r+3H(\rho_r+P_r)=0.
  \label{evoleqr}
\end{eqnarray}

\section{Specific $f(T)$ models}
\label{specificfTmodels}

In the previous section we reviewed the basic equations of $f(T)$ gravity and cosmology.
In this section we present some specific viable $f(T)$ models, and we provide the
formalism to quantify their deviation from $\Lambda$CDM cosmology in a unified way.

In order to  elaborate the modified Friedmann equations we introduce
\begin{eqnarray}
\label{Edef}
E^{2}(z)\equiv\frac{H^2(z)} {H^2_{0}}=\frac{T(z)}{T_{0}},
\end{eqnarray}
with $T_0\equiv-6H_{0}^{2}$. We mention that it proves more convenient to use as the
independent variable the redshift $z=\frac{a_0}{a}-1$, with $a_0$   the current scale
factor set to one for simplicity (in the following the subscript ``0'' marks the
current value of a quantity). Furthermore, assuming the matter to be dust, i.e.
$w_m\equiv P_m/\rho_m=0$, from (\ref{mattradevol}) we deduce that
$\rho_{m}=\rho_{m0}(1+z)^{3}$, and similarly imposing for the radiation
$w_r\equiv P_r/\rho_r=1/3$ from  (\ref{evoleqr}) we acquire
$\rho_{r}=\rho_{r0}(1+z)^{4}$.
Thus, the Friedmann equation (\ref{Fr11}) can be re-written as
\begin{eqnarray}
\label{Fr11EZ}
E^2(z,{\mathbf r})=\Omega_{m0}(1+z)^3+\Omega_{r0}(1+z)^4+\Omega_{F0} y(z,{\mathbf r})
\end{eqnarray}
with
\begin{equation}
\label{distortionpar}
 y(z,{\mathbf r})=\frac{1}{T_0\Omega_{F0}}\left[f-2Tf_T\right].
\end{equation}
In the above equations
we have defined
\begin{equation}
\label{OmF00}
\Omega_{F0}=1-\Omega_{m0}-\Omega_{r0} \;,
\end{equation}
with $\Omega_{i0}=\frac{8\pi G \rho_{i0}}{3H_0^2}$ the value of the corresponding density
parameter at present.
In summary, the effect of $f(T)$ gravitational modification is quantified by the function
 $y(z,{\mathbf r})$, that
is normalized to
unity at   present time, and which depends on $\Omega_{m0},\Omega_{r0}$, and on the
various parameters $r_1,r_2,...$ that are involved in the specific
$f(T)$ choice (assembled to form the vector ${\mathbf r}$).
Note that in the limit of $\Lambda$CDM
cosmology, namely when $f(T)=const.$, the function
$y(z,{\mathbf r})$ becomes a constant.

Let us now present all the %viable
specific $f(T)$ models that have been studied
in the literature, that include two parameters one of which is independent.
For each one of these models we calculate the
function $y(z,{\mathbf r})$ through (\ref{distortionpar}), and we quantify the
deviation of $y(z,{\mathbf r})$ from its  $\Lambda$CDM   constant value  using a
distortion
parameter $b$. The five %viable
$f(T)$ models are the following
\cite{Nesseris:2013jea}:

\begin{enumerate}
\item The power-law model \cite{Bengochea:2008gz}
(hereafter $f_{1}$CDM model), with
\begin{equation}
f(T)=\alpha (-T)^{b},
\label{power-lawmod}
\end{equation}
with $\alpha$ and $b$ the two parameters.
Inserting this $f(T)$ form
into  (\ref{Fr11}) at current time we
obtain
\begin{eqnarray}
\alpha=(6H_0^2)^{1-b}\frac{\Omega_{F0}}{2b-1},
\end{eqnarray}
while (\ref{distortionpar})
gives
\begin{equation}
\label{yLL}
y(z,b)=E^{2b}(z,b) \;.
\end{equation}
 Hence, when $b$ becomes zero the $f_{1}$CDM model
reduces to $\Lambda$CDM cosmology, namely
$T+f(T)=T-2\Lambda$, with $\Lambda=3\Omega_{F0}H_{0}^{2}$ and
$\Omega_{F0}=\Omega_{\Lambda 0}$).

\item The square-root exponential model (hereafter $f_{2}$CDM) \cite{Linder:2010py}
\begin{eqnarray}
f(T)=\alpha T_{0}(1-e^{-p\sqrt{T/T_{0}}}),
\label{Lindermod}
\end{eqnarray}
with $\alpha$ and $p$  the two parameters. In this case
(\ref{Fr11}) at present leads to
\begin{eqnarray}
\alpha=\frac{\Omega_{F0}}{1-(1+p)e^{-p}},
\end{eqnarray}
while (\ref{distortionpar}) gives
\begin{equation}
\label{yLLf2}
y(z,p)=\frac{1-(1+pE)e^{-pE}}{1-(1+p)e^{-p}}.
\end{equation}
 Since  $f_{2}$CDM reduces
to  $\Lambda$CDM cosmology  for $p \rightarrow +\infty$, we can
replace $p$ through $p=1/b$ obtaining
\begin{equation}
\label{yf2}
y(z,b)=\frac{1-(1+\frac{E}{b})e^{-E/b}}{1-(1+\frac{1}{b})e^{-1/b}},
\end{equation}
which tends to constant (unity) for
$b \rightarrow 0^{+}$.

\item  The   exponential model (hereafter $f_{3}$CDM) \cite{Nesseris:2013jea}:
\begin{eqnarray}
f(T)=\alpha T_{0}(1-e^{-pT/T_{0}}),
\label{f3cdmmodel}
\end{eqnarray}
where $\alpha$ and $p$ are the two model parameters.
In  this case
\begin{eqnarray}
\alpha=\frac{\Omega_{F0}}{1-(1+2p)e^{-p}},
\end{eqnarray}
while
\begin{equation}
\label{modfcdm3}
y(z,p)=\frac{1-(1+2pE^{2})e^{-pE^{2}}}{1-(1+2p)e^{-p}}.
\end{equation}
We can re-write  the above model using
 $p=1/b$, resulting to
\begin{equation}
\label{modfcdm3b}
y(z,b)=\frac{1-(1+\frac{2E^{2}}{b})e^{-E^{2}/b}}{1-(1+\frac{2}{b})e^{-1/b}},
\end{equation}
from which we can see that for $p \rightarrow +\infty$, or equivalently for
$b \rightarrow 0^{+}$, the $f_{3}$CDM model reduces to  $\Lambda$CDM cosmology.

\item The Bamba et al. logarithmic model
(hereafter $f_{4}$CDM) \cite{Bamba}
\begin{eqnarray}
f(T)=\alpha T_{0} \sqrt{\frac{T}{qT_{0}}}\;
{\text {ln}}\left( \frac{qT_{0}}{T}\right )
\label{f4cdmmodel}
\end{eqnarray}
with $\alpha$ and $q$  the two parameters.
Equation  (\ref{Fr11}) at present  time
gives
\begin{eqnarray}
\alpha=\frac{\Omega_{F0} \sqrt{q}}{2},
\end{eqnarray}
while (\ref{distortionpar})
yields
\begin{equation}
\label{modfcdm4}
y(z)=E(z) \;,
\end{equation}
Since the distortion function does not depend on the model
parameters, we can re-write (\ref{Fr11EZ}) as
\begin{eqnarray}
E(z)&=&\frac{1}{2}
\sqrt{\Omega^{2}_{F0}+4\left[\Omega_{m0}(1+z)^{3}+
\Omega_{r0}(1+z)^{4}\right]}+\frac{\Omega_{F0}}{2}.
\end{eqnarray}
We mention that  this model cannot reduce to $\Lambda$CDM
cosmology for any value of its parameters.

\item The hyperbolic-tangent model (hereafter $f_{5}$CDM) \cite{Wu:2011}:
\begin{eqnarray}
f(T)=\alpha(-T)^{n}{\text {tanh}}\left( \frac{T_{0}}{T}\right)
\label{f5cdmmodel}
\end{eqnarray}
with $\alpha$ and $n$  the two parameters.
In this case we acquire
\begin{eqnarray}
\alpha=-\frac{\Omega_{F0}(6H_{0})^{1-n}}{\left[ 2{\text {sech}}^{2}(1)+(1-2n){\text
{tanh}}(1)\right]},
\end{eqnarray}
and
{\small{
\begin{eqnarray}
y(z,n)=E^{2(n-1)}\frac{2{\text
{sech}}^{2}\left(\frac{1}{E^2}\right)+(1-2n)E^2{\text
{tanh}}\left(\frac{1}{E^2}\right)}{2{\text
{sech}}^{2}(1)+(1-2n){\text {tanh}}(1)}
\end{eqnarray}}}
 Similarly to the previous model, the  $f_{5}$CDM model cannot reduce to $\Lambda$CDM
cosmology for any value of
its parameters.

\end{enumerate}

In summary, the above five $f(T)$ models are the ones with up to two parameters,
out of which one is independent, that have been
studied in the literature \cite{Nesseris:2013jea}.
One could
definitely consider also their combinations, however the
appearance of many free
parameters is not a desirable feature. Thus, in the following we
investigate them separately.

\section{$f(T)$ models against cosmological data}
\label{observdata}

In this section we present the
observational data and the statistical methods that we use
in order to put constraints on the $f(T)$ models.
In particular, we use direct measurements of the
Hubble parameter, namely $H(z)$ data with the corresponding covariance
matrix, the
standard candles (SNIa and quasi-stellar objects: QSOs)
and finally the CMB shift parameter data.
Notice that in the case of the CMB shift parameter data,
we need to include the contribution of the radiation term $\Omega_{r0}$
in the normalized Hubble function. Here we utilize the following
formula $\Omega_{r0}=\Omega_{m0}a_{\text eq}$ \cite{Ywang16}, with
$a_{\text eq}=\frac{1}{1+2.5\times 10^{4}h^{2}(T_{\text CMB}/2.7K)^{-4}}$, where
we have set $T_{\text CMB}=2.7255K$ and $h=0.68$.

\subsection{$H(z)$ probes}
Let us start with %First, we utilize
the $H(z)$ Hubble data set as
compiled by Farooq \emph{et al.}, \cite{Farooq:2016zwm}.
This sample contains $N=38$ entries in
the following redshift interval
$0.07\leq z\leq 2.36$.
The novelty here is that we use for the first time (to our knowledge)
the covariance matrix of three BAO $H(z)$ measurements
\cite{Yu17,Yu17b} in constraining the $f(T)$ models.

Following standard lines, the nominal
chi-square function is written as
\begin{equation}
\label{eq:xtetr}
    \chi^2_{H}(\phi^{\mu}) = {\mathbf V} {\mathbf C}^{-1}_{\text cov} {\mathbf V}^{T},
\end{equation}
where $\phi^{\mu}$ is the statistical vector that contains
the free parameters,
${\mathbf C}^{-1}_{\text cov}$ is the inverse of the covariance matrix
\cite{Yu17,Yu17b} and
$$
    {\mathbf V} = \{H_{D}(z_1) - H_{M}(z_1,\phi^{\mu}),...,H_{D}(z_{N}) -
H_{M}(z_N,\phi^{\mu})\} \;.
$$
Also, $z_{i}$ are the observed redshifts, while the letters $M$ and $D$
denote the data and models respectively. In this context,
the theoretical Hubble parameter is parametrized as follows:
\begin{eqnarray}
\label{eq1}
H_{M}(z,\phi^{\mu}) = H_{0} E(z,\phi^{\mu+1})
\end{eqnarray}
and thus
{\small{
\begin{equation}
\label{eq:Vec}
    {\mathbf V} = \{H_{D}(z_1) - H_{0}E(z_1,\phi^{\mu+1}),.., H_{D }(z_{N}) -
H_{0}E(z_N,\phi^{\mu+1})\}
,
\end{equation}}}
where $H_{0}$ is the Hubble constant, $E(z)$ is the
dimensionless Hubble function
and the vector $\phi^{\mu+1}$
contains the other free parameters of the $f(T)$ models,
namely $(\Omega_{m0},b)$ etc.
%In this framework, the above vector ${\mathbf V}$ becomes

Clearly, using the usual $\chi^{2}$ estimator of Eq.(\ref{eq:xtetr})
in constraining the $f(T)$ gravity models we have to either
impose the exact value of $H_{0}$ or treating it as a free parameter.
The first option is rather inconvenient due to the
well known Hubble constant problem, namely
the observed Hubble constant ($H_{0} = 73.24 \pm 1.74$ Km/s/Mpc)
found by the SNIa team
(Riess et al. \cite{Riess16}) is in $\sim 3-3.5\sigma$ tension with that of
Planck (see $H_{0} = 67.8 \pm 0.9$ Km/s/Mpc; \cite{Planck16}).
On the other hand, if we treat
$H_{0}$ as a free parameter then we increase the parameter space.
Recently, \cite{Anagnostopoulos:2017iao} proposed a novel statistical
technique towards overcoming the above problems. Here we present
the main points of this method. Specifically,
inserting the vector (\ref{eq:Vec}) into Eq. (\ref{eq:xtetr})
we find after some simple calculations

\begin{eqnarray}
\chi^2(\phi^{\mu})=A H_{0}^2-2BH_{0}+\Gamma,
\end{eqnarray}
where
\begin{eqnarray}
&&A = {\mathbf E} {\mathbf C}^{-1}_{\text cov} {\mathbf E}^{T},
\nonumber\\
&&
B = \frac{1}{2} \left({\mathbf E} {\mathbf C}^{-1}_{\text cov}{\mathbf H}_{\text D}^T+
{\mathbf H}_{D} {\mathbf C}^{-1}_{\text cov}{\mathbf E}^T\right)\nonumber\\
&&
\Gamma = {\mathbf H}_{D}{\mathbf C}^{-1}_{\text cov}{\mathbf H}_{D}^T,\nonumber
\end{eqnarray}
with
$$
{\mathbf E}=\{E(z_1,\phi^{\mu+1}),...,E(z_N,\phi^{\mu+1}) \}
$$
and
$$
    {\mathbf H}_{D} = \{H_{D}(z_1),.., H_{D}(z_{N})\} \;.
$$

Therefore, the likelihood function of $\chi^{2}$, namely $\mathcal{L}=e^{-\chi^2/2}$, is
written as
%\begin{eqnarray}
% \mathcal{L}={\text exp}\left[\frac{A
%H_{0}^2-2BH_{0}+\Gamma}{2} \right]
%\end{eqnarray}
%or
\begin{eqnarray*}
\mathcal{L}(D|\phi^{\mu},M)=
{\text
exp}\left[\frac{A\left(H_{0}-\frac{B}{A}\right)^2-\frac{B^2}{A}+\Gamma}{2}\right]\;.
\end{eqnarray*}
Marginalizing over $H_{0}$ in the context of
Bayes's theorem we find
\begin{eqnarray}
p(\phi^{\mu}|D,M)=\frac{1}{p(D|M)}\int e^{-\frac{A(H_{0}-B/A)^2-B^2/A+\Gamma}{2}}dH_{0}.\,
\end{eqnarray}
Moreover, using the new variable $y=H_{0}-B/A$,
assuming that $H_{0} \in (0,+\infty)$ and introducing
flat priors $p(\phi^{\mu}|M,H_{0})= 1$, we arrive at
{\small{
\begin{equation}
p(\phi^{\mu}|D,M)=
\frac{1}{p(D|M)}  e^{ -\frac{1}{2} \left(\Gamma  -\frac{B^2}{A} \right)}
\sqrt{\frac{\pi}{2A}} \left[ 1 \!+ \! {\text erf} \left(  \frac{B}{\sqrt{2A} }\right)
\right],
\end{equation}}}
where ${\text erf}(x)=\frac{2}{\sqrt{\pi}}\int_{0}^{x}e^{-y^2}dy$ is
the error function.
To this end, it is trivial to show that the above
likelihood function reduces to a new ${\tilde \chi}^2_{H}$
function which is written as
\begin{equation}
\label{eq:marginalization}
{\tilde \chi}^{2}_{H}(\phi^{\mu+1})=
\Gamma  -\frac{B^2}{A}
+  \ln A
- 2 \ln \left[ 1 +  {\text erf} \left(  \frac{B}{\sqrt{2A} }\right) \right]\;.
\end{equation}
Notice that in the latter expression
the constant $ {\text ln}(\pi/2)$, is ignored since it
does not contribute in the minimization procedure.

Evidently, the estimator ${\tilde \chi}^2_{H}$ alleviates
the Hubble constant problem since it is not affected
by $H_{0}$. We would like to stress that
this is the first time that the current approach
is implemented towards testing the performance of the
$f(T)$ models against the $H(z)$ data.

\subsection{Joint analysis with other probes}

In order to place tight constraints on the model parameters
we use a joint likelihood analysis, involving the standard candles
(SNIa and QSOs, hereafter SC) and standard rulers (CMB shift parameter)
together with the $H(z)$ data. Notice, that standard candles and rulers
probe the integral of the Hubble parameter $H(z)$, implying that they give
indirect information of the cosmic expansion, while the $H(z)$ data provide
direct measurements of the expansion rate of the universe. Therefore, the
combination of the latter observational probes appears as an
ideal tool in constraining the $f(T)$ gravity models.

Specifically, we utilize the JLA sample of 740 SN Ia
of Betoule \emph{et al}. \cite{Betoule:2014frx}
%Specifically, we utilize the {\em Union 2.1} sample of 580 SN Ia of
%Suzuki \emph{et al}. \cite{Suzuki:12}
and the binned dataset of QSOs \cite{RisalitiLusso:2015,Roberts:2017} that
contains 24 entries.
It is interesting to mention that combining
the SNIa data with those of QSOs we manage to trace the
Hubble relation (distance modulus versus $z$) in the redshift
range $0.07< z< 6$.

Regarding the chi-square function of the standard candles
$\chi^{2}_{\text SC}$
we use the estimator of \cite{Nes05} which is not affected by the
value of the Hubble constant
(see also Ref.\cite{Chavez2016} and references therein), while for QSOs
we refer the reader the recent work of \cite{Roberts:2017}.
Moreover, we utilize the position of the CMB shift of acoustic peaks
provided by the Planck 2015 data points $(l_{a},R,z_{*})$.
The chi-square $\chi^{2}_{\text CMB}$,
the CMB shift parameter data, the theoretical formulas of
$(l_{a},R,z_{\star})$, the value of $\Omega_{r0}$ and the inverse of the
corresponding covariance matrix can be found in
Ref. \cite{Ywang16}.

Since the total likelihood function ${\cal L}_{\text tot}$ is defined as
the product of the individual likelihoods,
%${\cal L} \propto {\text exp}(-\chi^{2}/2)$,
namely $$
 \cal L_{\text tot}= {\cal L}_{H} \times {\cal L}_{\text SC} \times
{\cal L}_{\text CMB}   ,
$$
it is
implied that
$$
 \chi^2_{\text tot}={\tilde \chi}^2_{H}+\chi^2_{\text SC}+\chi^2_{\text CMB}\;.
$$

\begin{table*}[!]
\resizebox{1.0\textwidth}{!}
{
\begin{tabular}{ccccccccc} \hline \hline
Data & Model & $\Omega_{m0}$ & $b$ & $\alpha$ & $\beta$ & $\chi_{min}^{2}$ &${\text
AIC}$&
$|\Delta$AIC$|$ \vspace{0.05cm}\\ \hline
%----------------------------------------
                         &$f_{1}$CDM& $0.229 \pm 0.072 $ & $ 0.584 \pm 0.377 $ & - & -& $
18.967 $ &
 23.310 & 0.207 \vspace{0.01cm}\\
$H(z)$:                  &$f_{2}$CDM  &$0.284 \pm  0.029 $ & $0.600 \pm  0.410 $ & - & -
&
19.363 &
23.706 & 0.603 \vspace{0.01cm}\\
                         &$f_{3}$CDM& $ 0.297 \pm 0.077 $ & $0.266 \pm 0.169 $ & - & - &
20.741 &
25.084 & 1.981 \vspace{0.01cm}\\

                         &$\Lambda$CDM& $0.265 \pm 0.023 $ & - & - & - & 20.992 & 23.103
&
0.000 \vspace{0.45cm}\\ %%OK!
%------------------------------alls_good
%----------------------------------------
                         &$f_{1}$CDM& $0.249 \pm 0.029 $ & $0.258 \pm 0.253 $ & $0.141
\pm
0.058$ &
$3.102 \pm 0.635$ & 738.836 & 746.886 & 0.552 \vspace{0.01cm}\\ %%%%%OK!
$H(z)$/SC: 		         & $f_{2}$CDM & $0.263 \pm 0.021 $ & $0.301 \pm 0.148$ &
$0.141 \pm 0.059$ &
$3.102 \pm 0.650$ & 739.022& 747.072 & 0.738 \vspace{0.01cm}\\
                         &$f_{3}$CDM & $0.267 \pm 0.022 $ & $ 0.194\pm 0.069 $ & $0.141
\pm 0.069$ &
 $3.104 \pm 0.769$ & 739.441 & 747.491 & 1.157 \vspace{0.01cm}\\ %%%%OK!

			             &$\Lambda$CDM& $0.269 \pm 0.021 $ & - & $0.142 \pm
0.041$ & $3.112 \pm 0.474$ & 740.
304 & 746.334 & 0.000 \vspace{0.45cm}\\  %%%%%NAI
%-----------------------------------alls_good
%----------------------------------------% % % % % % % % % % % % %
                         &$f_{1}$CDM& $  0.307 \pm 0.003 $ & $ -0.018 \pm 0.029 $ &
$0.141
\pm 0.
033$ & $3.101 \pm 0.414$ & 743.964 & 752.014 & 1.546 \vspace{0.01cm}\\

$H(z)$/SC/CMB$_{\text shift}$: & $f_{2}$CDM& $ 0.305\pm 0.001 $ & $  0.047 \pm 0.082 $ &
$0.141 \pm 0.
033$ & $3.101 \pm 0.412$ & 744.438 & 752.488 & 2.020 \vspace{0.01cm}\\
                         &$f_{3}$CDM& $0.305 \pm 0.001$ & $ 0.050 \pm 0.061$& $0.142 \pm
0.032$ &
$3.100 \pm 0.203$ & 744.446 & 752.496 & 2.028 \vspace{0.01cm}\\

                         &$\Lambda$CDM& $0.305 \pm 0.001 $ & - & $0.141 \pm 0.033$ &
$3.101 \pm 0.
412$ & 744.438 & 750.468 & 0.000 \vspace{0.45cm}\\  %%OK!
%----------------------------------------
\hline\hline
\end{tabular}}
\caption[]{Cosmological constraints. The first column indicates the data set (s) used, the 
second 
column includes the cosmological models used in this study the third and fourth columns 
provide the 
$\Omega_{m0}$ and $b$ best fit values. The next two columns show the values of the 
JLA-data intrinsic free parameters, namely $\alpha$ and $\beta$. The last three columns 
present the 
goodness-of-fit 
statistics ($\chi^{2}_{min}$, AIC and $\Delta${\text AIC}$={\text AIC}_{\text 
model}-{\text AIC}_{\text min}|$). The abbreviation SC denotes standard candles, namely 
SNIa and QSOs.\label{tab:growth1}
}
\end{table*}

\begin{table*}[!]
\resizebox{1.0\textwidth}{!}
{
\begin{tabular}{cccccccccc} \hline \hline
Data & Model & $\Omega_{m0}$ & $b$ & $\alpha$ & $\beta$ & $h$ &$\chi_{min}^{2}$ &${\text 
AIC}$& $|\Delta$AIC$|$ \vspace{0.05cm}\\ \hline
%----------------------------------------
        &$f_{1}$CDM& $0.226 \pm 0.077 $ & $ 0.603 \pm 0.374 $ & - & -& $0.651\pm0.037$& 
$19.768 $ & 
26.474 & 0.205 \vspace{0.01cm}\\        
$H(z)$: &$f_{2}$CDM  &$0.285 \pm  0.029 $ & $0.622 \pm  0.426 $ & - & -& $0.644\pm0.039$ 
&20.145 &26.851 & 0.582 \vspace{0.01cm}\\
        &$f_{3}$CDM& $ 0.310 \pm 0.108 $ & $0.292 \pm 0.199 $ & - & - & $0.639\pm0.115$ 
&21.529 &28.235 & 1.966 \vspace{0.01cm}\\
        &$\Lambda$CDM& $0.266 \pm 0.018 $ & - & - & - &$0.696\pm0.014$& 21.926 & 26.269 
&0.000 \vspace{0.45cm}\\ %%OK!
%------------------------------alls_good%----------------------------------------
           &$f_{1}$CDM& $0.249 \pm 0.028 $ & $0.260 \pm 0.234 $ & $0.141\pm0.055$ &$3.103 
\pm 0.614$ &$0.686\pm0.020$& 739.740 & 749.815 & 0.534 \vspace{0.01cm}\\ %%%%%OK!
           $H(z)$/SC: & $f_{2}$CDM & $0.264 \pm 0.021 $ & $0.302 \pm 0.134$ &$0.141 \pm 
0.055$ &$3.103 \pm 0.621$ & $0.686\pm0.020$ & 739.926& 750.001 & 0.720 \vspace{0.01cm}\\
           &$f_{3}$CDM & $0.267 \pm 0.022 $ & $ 0.194\pm 0.066 $ & $0.141\pm 0.065$ 
&$3.104 \pm 0.735$ &$0.687\pm0.021$ & 740.351 & 750.426 & 1.145 \vspace{0.01cm}\\ %%%%OK!
              &$\Lambda$CDM& $0.269 \pm 0.021$ & - & $0.142 \pm0.041$ & 
$3.112 \pm 0.474$ & $0.694\pm0.016$ &
 741.231 & 749.281 & 0.000 \vspace{0.45cm}\\  %%%%%NAI
%-----------------------------------alls_good%----------------------------------------% % 
% % % % % 
% % % % % %
                               &$f_{1}$CDM& $  0.307 \pm 0.003 $ & $ -0.020 \pm 0.028 $ 
&$0.142\pm 0.007$ & $3.101 \pm 0.080$ &$0.671\pm0.007$& 744.804 & 754.879 & 
1.466\vspace{0.01cm}\\
$H(z)$/SC/CMB$_{\text shift}$: & $f_{2}$CDM& $ 0.305\pm 0.001 $ & $  0.040 \pm 0.080 $ 
&$0.141 \pm 0.006$ & $3.100 \pm 0.080$ & $0.671\pm0.007$ & 745.357 & 755.432 & 2.019 
\vspace{0.01cm}\\        &$f_{3}$CDM& $0.305 \pm 0.001$ & $ 0.054 \pm 0.072$& $0.142 
\pm0.007$ &$3.097 \pm 0.080$ & $0.671\pm0.007$ & 745.365 & 755.440 & 2.027 
\vspace{0.01cm}\\          &$\Lambda$CDM& $0.305 \pm 0.001 $ & - & $0.142 \pm 0.007$ 
&$3.100 \pm 0.080$ & $0.671\pm0.007$ & 745.363 & 753.413 & 0.000 \vspace{0.45cm}\\  %%OK!
%----------------------------------------
\hline\hline
\end{tabular}}
\caption[]{Same as Table \ref{tab:growth1}, however now the parameter 
$h\equiv\frac{H_0}{100 \textrm{km} \textrm{s}^{-1} \textrm{Mpc}^{-1}}$ is not marginalized 
over, but it is free to vary.\label{tab:growth2}}
\end{table*}

In order to test the statistical significance of our
constraints we implement the AIC \citep{Akaike1974}
criterion.
%and BIC \citep{bic} criteria respectively.
In particular, considering Gaussian errors the
corresponding estimator is
\begin{eqnarray}
{\text AIC} = -2 \ln {\cal L}_{\text max}+2k+\frac{2k(k+1)}{N-k-1} \label{eq:AIC}\;,\\
%{\text BIC} = -2 \ln {\cal L}_{\text max}+k\ln{N}\;.
\end{eqnarray}
where $N$ and $k$ denote the total number of data and
the number of fitted parameters (see also \citep{Liddle:2007fy}).
Obviously, a smaller value of AIC means a better model-data fit.
In this framework, if we want to test
the performance of the different cosmological models in fitting
the observational data then we need to introduce the model
pair difference, namely
$\Delta {\text  AIC}={\text AIC}_{\text model}-{\text AIC}_{\text min}$.
Therefore, the inequalities
$4<\Delta {\text AIC} <7$ indicate a positive evidence
against
the model with higher value of ${\text AIC}_{\text model}$ \cite{Ann2002, Burham2004},
while the
restriction $\Delta {\text AIC} \ge 10$ suggests a strong such evidence.
On the other hand, the condition $\Delta {\text AIC} \le 2$ implies an indication
of consistency between the two comparison models.

In summary, an overall presentation of our constraints is listed in Table
\ref{tab:growth1} for various cosmological probes. The Table
contains the fitted model parameters, including the
intrinsic values of JLA $(\alpha,\beta)$,
and the goodness of fit statistics ($\chi_{\text min}^2$, AIC), for the
specific $f(T)$ gravity models.
For comparison we additionally provide the results of the
usual $\Lambda$CDM cosmological model.  Furthermore, in 
Table \ref{tab:growth2} we show the results after the parameter $H_0$ has been allowed to 
vary. 
Comparing the two Tables we see that our results and the two methods of analysis are 
perfectly 
consistent with each other. 
Furthermore, from Table \ref{tab:growth2} 
we observe that the $f(T)$ models support a smaller value of 
Hubble constant than the value obtained from Cepheids. 
In particular, for all $f(T)$ models we find 
$H_{0}=67.1 \pm 0.7$Km/s/Mpc 
which is closer to the Planck value.

Below, we provide a qualitative discussion of our constraints, giving the reader the
opportunity to appreciate the new results of our study. Notice that we have excluded
models $f_{4}$CDM and $f_{5} $CDM from the rest of the analysis, since we have confirmed
the results of Nesseris et al. \cite{Nesseris:2013jea}, namely that the observational
data disfavor the above two $f(T)$ models at high significance level.

\subsubsection{Fitting $f(T)$ models with $H(z)$ data}

In the light of the new chi-square estimator
[see Eq.(\ref{eq:marginalization})] we utilize
for the first time
the covariance matrix of $H(z)$ data
\cite{Yu17,Yu17b} in order to constrain the $f(T)$ models.
Let us now briefly present our results.
For comparison we also
provide the results of Nunes et al. \cite{Nun16}.

Specifically, we find:

\begin{itemize}

\item For $f_{1}$CDM model:
${\tilde \chi}^2_{H}=18.967$ (AIC=23.310),  $\Omega_{m0}=
0.229 \pm 0.072$ and $b = 0.584  \pm 0.377 $, while
\cite{Nun16} found:
$(\Omega_{m0},b)=
(0.231_{-0.019}^{+0.016},0.033_{-0.035}^{+0.045})$ for $H_{0}=72.85_{-1.8}^{+1.7}$
Km/s/Mpc.

\item For $f_{2}$CDM model:
${\tilde \chi}^2_{H}=19.363$ (AIC=23.706),
$\Omega_{m0}=0.284 \pm 0.029$ and $b = 0.600 \pm 0.410$.
Notice that \cite{Nun16} obtained:
$(\Omega_{m0},b)=
(0.243_{-0.015}^{+0.014},0.112_{-0.035}^{+0.045})$ for $H_{0}=71.53 \pm 1.3$ Km/s/Mpc.

\item For $f_{3}$CDM model:
${\tilde \chi}^2_{H}= 20.741$ (AIC=25.084),  $\Omega_{m0}=
0.297 \pm 0.077$ and $b = 0.266 \pm 0.169$.
To this end, \cite{Nun16} found:
$(\Omega_{m0},b)=
(0.242_{-0.015}^{+0.013},0.106_{-0.090}^{+0.052})$ for $H_{0}=71.57 \pm 1.3$ Km/s/Mpc.

\item For $\Lambda$CDM model:
${\tilde \chi}^2_{H}=20.992$ (AIC=23.103)
and $\Omega_{m0}=0.265 \pm 0.023$.

\end{itemize}

\begin{figure*}[t]
\centering
\includegraphics[width=0.9\textwidth]{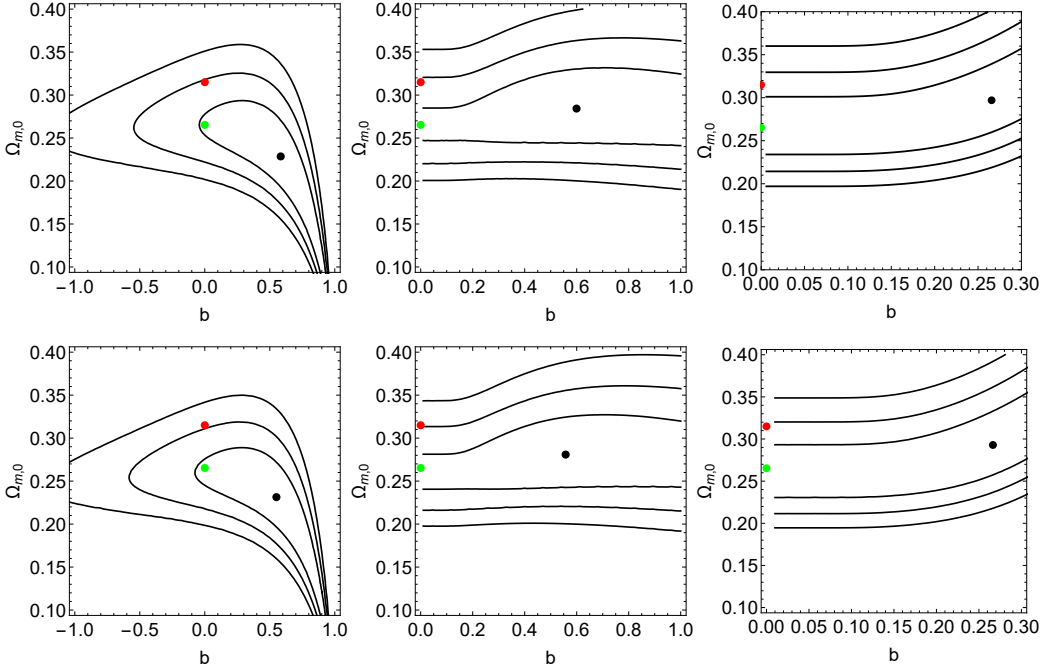}
\caption{Likelihood contours in the $b-\Omega_{m0}$ plane using
the $H(z)$ data. The
contours correspond to $1\sigma$, $2\sigma$ and $3\sigma$
confidence levels.
From left to right, observational
constraints for power-law $f_1$CDM
(\ref{power-lawmod}), $f_2$CDM model (\ref{Lindermod}) and $f_3$CDM model
(\ref{f3cdmmodel}). From bottom to top, constraints using $H(z)$
data without covariance and adding covariance. The black point
represents the best fit solutions for the current $f(T)$ models (see Table I).
Notice, that the red (or green) dot point corresponds to the flat
$\Lambda$CDM model using the Planck \cite{Planck16}
[or $H(z)$] best fit solution.
}
\label{fig:Hz_comparison}
\end{figure*}

In Fig. 1 we present the 1$\sigma$, 2$\sigma$ and $3\sigma$
contours in the $(b,\Omega_{m0})$ plane
for the $f_{1-3}$CDM models.
Taking into account the best-${\tilde \chi}^{2}$
and the value of AIC we find that the best model
is the $f_1$CDM, while
there is a mild tension between the $f_3$CDM models
and the $H(z)$ data $\Delta$AIC$={\text AIC}_{f_1}-{\text AIC}_{f_3} > 2$.
Moreover, the fact $\Delta$AIC$\le 2$ implies that the
$f_{1}$CDM model is statistically equivalent with $\Lambda$CDM and $f_2$CDM.

Lastly, we would like to stress that the
aforementioned $\Omega_{m0}$ constraints are in agreement
within $1\sigma$ errors with those of
Nunes et al. \cite{Nun16} who considered the case
where the covariance matrix of the $H(z)$ data is diagonal.
However, our results regarding the $b$ parameters of $f_{1,2}$CDM
models are somewhat larger ($\sim 1.2-1.5 \sigma$) with
those of \cite{Nun16}, while in the case of
$f_{3}$CDM model the constraints
are similar (within $1\sigma$) in both studies.
We mention that in the work of Nunes et al. \cite{Nun16} the
cosmic chronometer data, which are based on the
relative ages of the passively evolving galaxies, were used.
This sample contains the 30 $H(z)$ measurements
in the redshift range $0<z<2$.
Moreover, Nunes et al. \cite{Nun16}
used the standard estimator (\ref{eq:xtetr}) and
thus they had to treat the
value of $H_{0}$ as a free parameter.
In order to understand the effectiveness of the $H(z)$
covariance matrix and the new chi-square estimator in constraining
the $b-\Omega_{m0}$ solution space, we compare the $b-\Omega_{m0}$ contours with
and without covariance.

\subsubsection{Fitting $f(T)$ models with $H(z)$, standard candles and rulers}

Firstly, we combine the observed Hubble relation provided
by SNIa and QSOs data and the direct measurements
of the Hubble parameter. The results are as follows.
Regarding the $f(T)$
models we obtain
$(\Omega_{m0},b)=(0.249 \pm 0.029,0.258 \pm 0.253)$,
$(\Omega_{m0},b)=(0.263 \pm 0.021,0.301 \pm 0.148)$ and $(\Omega_{m0},b)=(0.267 \pm
0.022,0.194\pm 0.
069)$ for the $f_1$CDM, $f_2$CDM and $f_3$CDM models, respectively, with $\chi^{2}_{\text
min}= 738.
836 $, $\chi^{2}_{\text min}= 739.022 $ and $\chi^{2}_{\text min}= 739.441$. The latter
constraints are
in
agreement within $1\sigma$ (within $1\sigma$ uncertainties)
with those of Nunes et al. \cite{Nun16} who found, combing
cosmic chronometer, SNIa (Union 2.1) and BAO data,
$(\Omega_{m0},b)=(0.234^{+0.016}_{-0.014},0.051^{+0.025}_{-0.019})$
with $H_{0}=72.75^{+1.7}_{-1.8}$Km/s/Mpc for
$f_1$CDM,
$(\Omega_{m0},b)=(0.278^{+0.01}_{-0.02},0.133^{+0.043}_{-0.130})$
with $H_{0}=68.19^{+1.90}_{-0.93}$Km/s/Mpc for
$f_2$CDM and
$(\Omega_{m0},b)=(0.266^{+0.009}_{-0.010},0.09^{+0.041}_{-0.08})$
with $H_{0}=69.8^{+0.89}_{-0.84}$Km/s/Mpc for
$f_3$CDM models respectively.

%Including the CMB shift parameters data
%in the likelihood analysis we find
%$(\Omega_{m0},b)=(0.307 \pm 0.002,-0.018 \pm 0.027 )$,
%$(\Omega_{m0},b)=(0.305 \pm 0.001,0.075 \pm 0.086)$
%and $(\Omega_{m0},b)=(0.305\pm 0.001,0.045\pm 0.075)$
%for $f_1$CDM, $f_2$CDM and $f_3$CDM
%models, respectively, with $\chi^{2}_{\text min}$ of $\sim 751-752$.

Including the CMB shift parameters data in the
likelihood analysis we find
$(\Omega_{m0},b)=(0.307 \pm 0.003,-0.018 \pm 0.029)$,
$(\Omega_{m0},b)=(0.305 \pm 0.001,0.047 \pm 0.082)$
and $(\Omega_{m0},b)=(0.305\pm 0.001,0.050\pm 0.061)$
for the $f_1$CDM, $f_2$CDM and
$f_3$CDM models, respectively, with $\chi^{2}_{\text min}$ of $\sim 744$.
Evidently, using the joint cases we find that
the best model is $\Lambda$CDM.
Based on AIC we observe that the  $f_1,f_2,f_3$ models
fit at a statistically acceptable level either the
$H(z)$ or $H(z)$/SNIa/QSOs data.

%Based on AIC we observe that the  $f_{1-3}$CDM
%models fit at a statistically
%acceptable level   the $H(z)$/SNIa/QSOs data.

%or $H(z)$/SNIa/QSOs/CMB$_{\text shift}$ data.
In contrast, there is a weak evidence ($\Delta$AIC$\sim 2$)
against $f_{2,3}$CDM from the $H(z)$/SNIa/QSOs/ CMB$_{\text shift}$ data, while
the fact $\Delta$AIC$<2$ implies
that the power-law $f_{1}$CDM model is statistically equivalent with
that of $\Lambda$CDM.
Furthermore, we would like to point that our results are
in agreement (within $1\sigma$ errors) with those of
Nesseris et al. \cite{Nesseris:2013jea}
who used a combination of SNIa/BAOs/CMB$_{\text shift}$ data.
Concerning the $f_{1}$CDM model the
present results can be compared with those
of Nunes (2018) \cite{Nunes:2018xbm}
who used the CMB power spectrum. Lastly, our constraints
for $f_{1}$CDM and $f_{2}$CDM models
are in very good agreement with those of
and Xu et al. \cite{Xu:2018npu} who combined
SNIa/$H(z)$BAO/CMB$_{\text shift}$ data.
Concerning $f_{3}$CDM model our best fit value for $b$
is somewhat lower with that of \cite{Xu:2018npu}.
Specifically, these authors found: (a)
$H_{0}=69.4\pm 0.8$Km/s/Mpc, $\Omega_{m0}=0.298\pm 0.007$
and $b=-0.10^{+0.09}_{-0.07}$ for $f_{1}$CDM model,
(b)
$H_{0}=69.6\pm 0.9$Km/s/Mpc, $\Omega_{m0}=0.296\pm 0.007$
and $b=0.13^{+0.09}_{-0.11}$ for $f_{2}$CDM model,
and (c)
$H_{0}=69.5\pm 0.8$Km/s/Mpc, $\Omega_{m0}=0.297\pm 0.007$
and $b=0.41\pm 0.31$ in the case of $f_{3}$CDM model.

%Specifically he found: (a)
%$H_{0}=72.65^{+3.04}_{-4.35}$Km/s/Mpc, $b=0.0045^{+0.0037}_{-0.0041}$,
%$10^{2}\Omega_{b0}h^{2}=2.235^{0.015}_{-0.014}$,
%$\Omega_{\text cdm0}h^{2}=0.1181\pm 0.001$ for CMB$_{\text spectrum}$/BAO
%(b) $H_{0}=73.47^{+2.3}_{-2.1}$Km/s/Mpc, $b=0.0053^{+0.0023}_{-0.0020}$,
%$10^{2}\Omega_{b0}h^{2}=2.234 \pm 0.014$,
%$\Omega_{\text cdm0}h^{2}=0.118\pm 0.0011$ for CMB$_{\text spectrum}$/BAO/$H_{0}$.

\begin{figure*}[!t]
\includegraphics[width=0.99\textwidth]{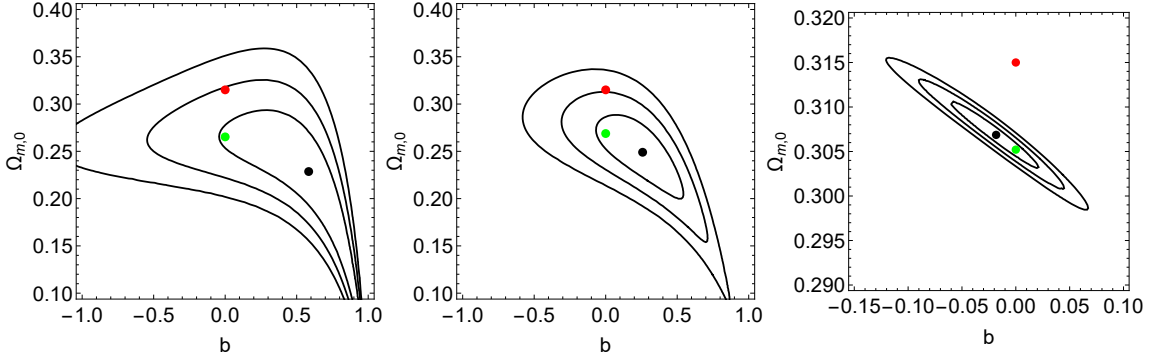}
\caption{Likelihood contours in the case
of the power-law $f_1$CDM model,
using (left to right) the following data:
(a) $H(z)$ data, (b) $H(z)$/QSO/SNIa and (c) $H(z)$/QSO/SNIa/CMB$_{\text shift}$.
The red point corresponds
to the
Planck $\Lambda$CDM model \cite{Planck16}, while our best
fit solutions are given by
green ($\Lambda$CDM) and black ($f_1$CDM) points respectively.
}
\label{fig:f1_3pack}
\end{figure*}

\begin{figure*}[!t]
\includegraphics[width=0.99\textwidth]{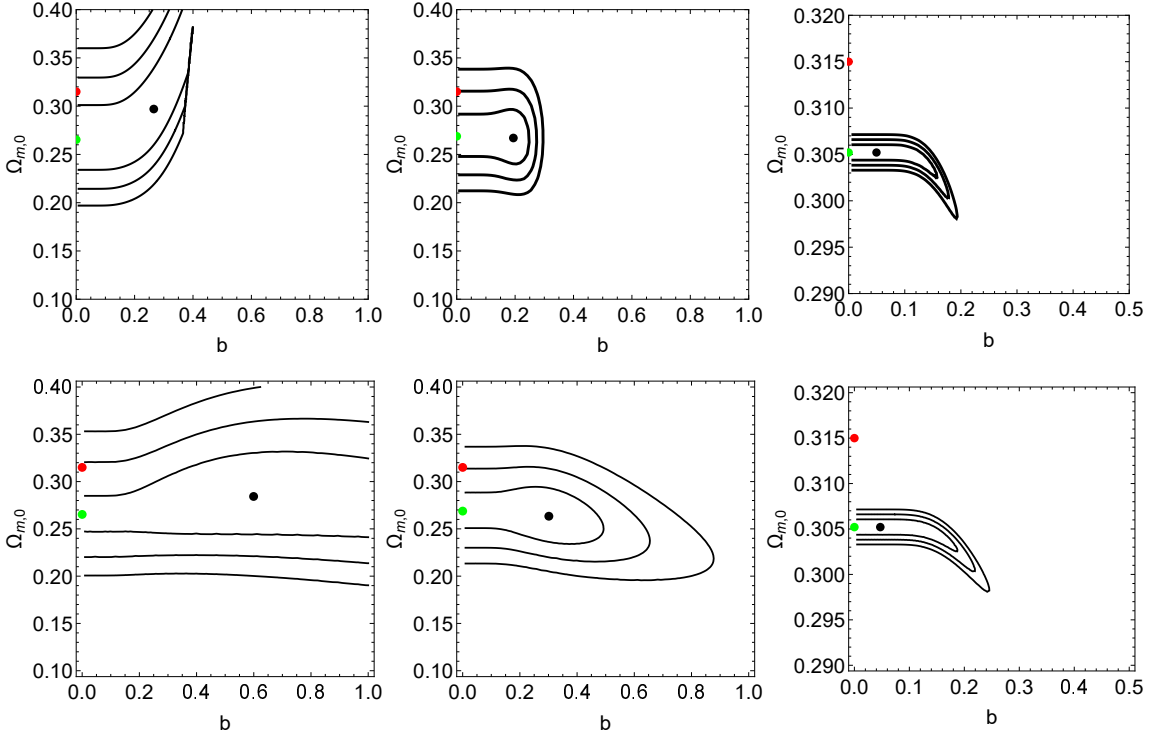}
\caption{Likelihood contours in the case of $f_2$CDM (bottom panel)
and $f_3$CDM (top panel) models.
From left to right we have
(a) $H(z)$ data, (b) $H(z)$/QSO/SNIa and (c) $H(z)$/QSO/SNIa/CMB$_{\text shift}$.
The red point corresponds to the
Planck $\Lambda$ CDM model \cite{Planck16}, while our best fit solutions
are given by
green ($\Lambda$CDM) and black ($f_{2,3}$CDM) points respectively.
}
\label{fig:joint_HzSNIa}
\end{figure*}

Notice, in Figs. \ref{fig:f1_3pack} and \ref{fig:joint_HzSNIa} we
show the corresponding confidence contours.
Comparing the first panel with the other two, we see that the  Figure-of-Merit
(FoM)\footnote{The FoM is defined as the inverse of the enclosed area of the 2$\sigma$
contour in the solution space of any two degenerate free parameters, is our case
$b-\Omega_{m0}$.} of $f(T)$ models is increased by a factor of $\sim 2.5$
in the case of $H(z)$/SNIa/QSOs, while for $H(z)$/SNIa/QSOs/CMB$_{\text shift}$
we find a two-fold increase of FoM.

Finally, in order to give the reader 
the opportunity to appreciate the new results of our study
we conclude this section with a brief discussion regarding  
our new and novel statistical results.

Although, 
our observational constraints are in qualitative agreement
with previous studies on  
$f(T)$ gravity 
\cite{Wu:2010mn,Bengochea001,Cardone:2012xq,Nesseris:2013jea,Basilakos:2016xob,
Nunes:2016plz,Capozziello:2015rda,Nunes:2018xbm,Xu:2018npu},
we would like to spell out clearly the reasons 
of which our analysis improves the observational
constraints of $f(T)$ models with respect to
previous studies. Firstly, 
we include for the first time the
covariance matrix of $H(z)$ data in $f(T)$ models,
and we utilize 
at the same time %and we utilize
the recently proposed statistical
method of \cite{Anagnostopoulos:2017iao}, which is not affected 
by the Hubble constant problem. Secondly,
%(b)
we use the total amount of 
standard candles, namely we  
combine the SNIa data with those of QSOs and thus
we trace the Hubble relation in the redshift
range $0.07< z< 6$.


%%%%%%%%%%%%%%%%%%%%%%%%%%%%%%%%%%%%%%%%%%%%%%%%%%%%%%%%%%%%%%%%%%%%%%%%%%
%%%%%%%%%%%%%%%%%%%%%%%%% Monte Carlo simulations %%%%%%%%%%%%%%%%%%%%%%%
%%%%%%%%%%%%%%%%%%%%%%%%%%%%%%%%%%%%%%%%%%%%%%%%%%%%%%%%%%%%%%%%%%%%%%%%%%

\section{Constraining $f(T)$ models with future $H(z)$ data}
As we have already discussed the statistical analysis of
section 4.B.1 indicates that direct measurements of the Hubble
expansion favor the power law $f(T)$ model over the other
models, including that of $\Lambda$CDM.
In this section we investigate the impact of
using future Hubble parameter data, based on the next generation of surveys,
%to constrain the $f(T)$ gravity models.
to distinguish the power law $f(T)$ model from the expectations
of $\Lambda$CDM.

In particular, we are interested to check
how better can we go in constraining the
$f(T)$ models by increasing the number of
the present $H(z)$ measurements from 38 to 100.
A detailed discussion concerning this test %the method
can been found in \cite{Anagnostopoulos:2017iao}, where the authors applied
their algorithm on the scalar-field dark energy models.
Specifically, we generate a large sets of Monte Carlo simulations
towards quantifying the
ability of future $H(z)$ data to place strong constraints
on the $f(T)$ models. Notice that this is the first time
in the literature that such an analysis is applied on the
$f(T)$ gravity models.
The algorithm is developed via a three-step process, a brief
description of which is as follows.

\begin{itemize}

\item Firstly, we select the viable $f_1$CDM model with
$(\Omega_{m0},b,H_{0})=(0.214,0.653,68)$ as a reference model.

\item Secondly, we peak a redshift $z_{\text ran} \in [0.07, 2.36]$,
by randomly sampling the redshift distribution of the current $H(z)$ sample,
and we extract the measured Hubble parameter $H_{D}(z_{\text ran})$
as well as the ideal Hubble function $H_{\text ref}(z_{\text ran})$
from the reference $f_1$CDM model. Then we
  derive the
deviation of the observed Hubble parameter from the
fiducial $f_1$CDM model by randomly sampling
the distribution of the %absolute
differences $\delta H=|H_{D}-H_{\text ref}|$.

\item In order to produce the mock $H(z)$ data\footnote{We sample
the number of mock data   $N\in [38,120]$ in steps of 10.}
the corresponding mock Hubble parameter $H_{\text MC}$ is chosen
from the following normal distribution
${\cal N}(H_{\text ref},\sigma^{2}_{\text ran})$, where
$\sigma_{\text ran}=\sqrt{\sigma_{H}^{2}+\delta H^{2}}$.
Therefore, for each mock entry we provide the
following simulated triad
$ \{ z_{\text ran},H_{\text MC},\sigma_{\text ran} \}_{j}$, where $j=1,..N$ and
$N\in [38,120]$.

\end{itemize}

\begin{figure}[ht]
\begin{center}
\includegraphics[width=0.5\textwidth]{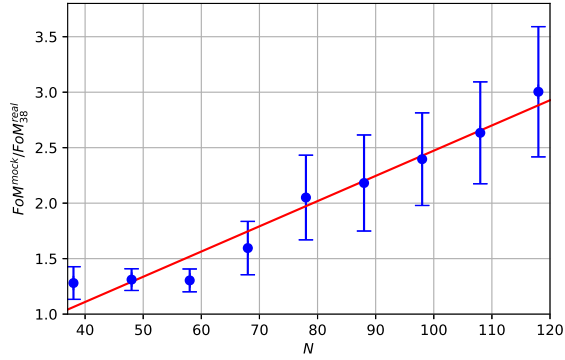}
\caption{ The $FoM/FoM_{38}$ as a function of the number of
entries in the mock $H(z)$ sample.
The quantity $FoM_{38}$ is the Figure-of-Merit of the
current $H(z)$ data. The reference model is
the power-law $f_1$CDM model (\ref{power-lawmod}) with
$(\Omega_{m0},b,H_{0})=(0.259,0.121, 68.603)$.
}
\label{fig:simulations}
\end{center}
\end{figure}

\begin{figure}[ht]
\begin{center}
\includegraphics[width=0.6\textwidth]{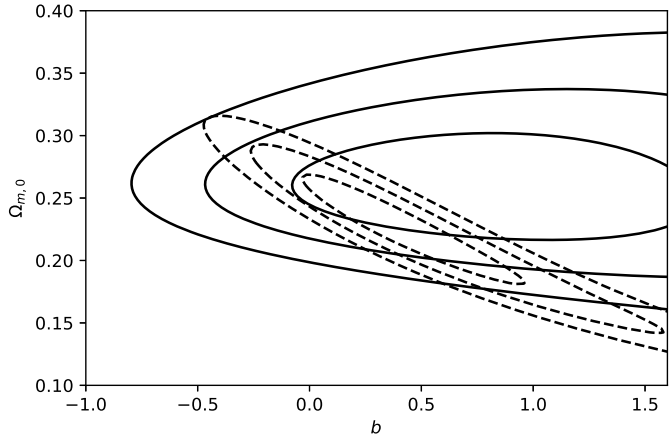}
\caption{ The $1\sigma$, $2\sigma$ and $3\sigma$ likelihood
contours in the $b - \Omega_{m0} $ plane resulting from
likelihood analysis of the real $H(z)$ data set (black lines)
and the mock $H(z)$ sample with $N = 100$ entries (red contours).
The reference model is the power-law $f_1$CDM model
(\ref{power-lawmod}) .  }
\label{fig:BeforeAndAfter}
\end{center}
\end{figure}

In Fig. \ref{fig:simulations} we present the ratio FoM/FoM$_{38}$
versus $N$, for which we have performed 100 Monte-Carlo realizations for
each chosen number ($N=38,40,..120$) of the simulated $H(z)$ data.
Notice that FoM$_{38}$ is the Figure-of-Merit of the observed
$H(z)$ data. Using standard linear regression we obtain the following
relation:
\begin{equation}
%    FoM =  (0.036\pm 0.004) N + 0.045 \pm  0.351.
    \frac{FoM}{FoM_{38}} =  (0.023\pm 0.0001) N + 0.200 \pm  0.018.
\end{equation}
Based on these new forecasts we argue that it is realistic
to expect that a future sample of $\sim 100-120$ $H(z)$ measurements
will increase the present FoM of the $f_{1}$CDM model by a factor
of $\sim 2.5-3$.  In order to visualize this improvement
 in Fig. \ref{fig:BeforeAndAfter} we present
the contours of one simulation of 100 $H(z)$ measurements
in the $b-\Omega_{m0}$ plane (red-scale contours). On top on that
we show the corresponding
 contours resulting from the current $H(z)$ sample.
The improvement is obvious.
Indeed, our Monte Carlo analysis indicates that a
future sample of $\sim 100$ $H(z)$ measurements
in the redshift interval $0<z<2.4$, will be an
ideal and indispensable tool towards testing the
viability of the $f(T)$ gravity models.

\section{Conclusions}
\label{Conclusions}
In this work we have extracted observational
constraints on the viable $f(T)$ gravity models, using
the new $H(z)$ data together with the
corresponding covariance matrix and
the recently proposed statistical
method of \cite{Anagnostopoulos:2017iao}, which is
not affected by the value of $H_0$ and
thus it bypasses the problem of the disagreement
in its exact numerical value between Planck and SNIa measurements.
In particular, one
marginalizes analytically the Hubble constant in the likelihood
function using Bayesian statistics, which allows to circumvent the
$H_0$ value problem and thus to reduce the parameter space
without adding numerical complexity.
Hence, the aforementioned approach leads to more robust results
with respect to previous studies.
Finally, apart from the $H(z)$ data, we also proceeded to joint
analysis using
standard candles such as SNIa and
quasi-stellar objects (QSOs), as
well as CMB shift parameters data from Planck.
To our knowledge,
this is the first time that
the $H_0$-independent method of the $H(z)$ data,
the covariance matrix of the $H(z)$ measurements, the JLA and the QSOs data
are combined with the Planck CMB shift parameter
in constraining the $f(T)$ models.

We considered five $f(T)$ models, each with two
independent parameters, with three
of these models being viable since they pass the basic observational tests.
In all of them we quantified their deviation from $\Lambda$CDM cosmology
through a sole parameter.
Hence, we used the aforementioned observational
data in order to fit the model parameters.
Furthermore, we applied the AIC criterion
in order to compare our results
with those of $\Lambda$CDM cosmology.

As we saw, the incorporation of more data sets
through joint analyses substantially improves the
fitting features for all models, and practically all
three examined $f(T)$ models, are
very efficient and in very good agreement with observations.
Among them, $f_{1}$CDM
model, namely the power-law  $f(T)$ gravity, is the one with the best
fitting behavior, and the one
in which a small but non-zero deviation from
$\Lambda$CDM cosmology is slightly favored (the deviation
parameter is non-zero at 1$\sigma$ confidence level).  Nevertheless, the
corresponding AIC
value reveals that $f_{1}$CDM
model is statistically equivalent with $\Lambda$CDM paradigm.
On the other hand, for $f_{2}$CDM
and $f_{3}$CDM scenarios, deviation from $\Lambda$CDM cosmology
is also allowed,
however the best-fit values are very close to their $\Lambda$CDM one.

Our results are in qualitative agreement
with previous observational investigations on
$f(T)$ gravity, where the SNIa/BAO data
alongside the CMB (shift parameter or power spectrum)
data had been used
\cite{Wu:2010mn,Bengochea001,Cardone:2012xq,Nesseris:2013jea,Basilakos:2016xob,
Nunes:2016plz,Capozziello:2015rda,Nunes:2018xbm,Xu:2018npu}.
However, our analysis improves further the observational
constraints with respect to
previous studies, since (a) we introduce for the first time the
covariance matrix of $H(z)$ data in $f(T)$ models and we use
the recently proposed statistical
method of \cite{Anagnostopoulos:2017iao}, which is
independent from the Hubble constant problem, and (b)
we combine the SNIa data with those of QSOs and thus
we manage to trace the Hubble relation in the redshift
range $0.07< z< 6$. 
 However, the measurements of the 
distance modulus provided by the QSO data suffer from relatively large errors. 

Now, regarding the importance of using 
direct measurements of the 
Hubble expansion some considerations are in order at this point.
In general, the $H(z)$ data
are the only data
which are providing a direct measurement of the Hubble expansion 
as a function of redshift. Obviously, this feature makes 
them ideal tools for studying 
the accelerated expansion of the universe.

It is well known that the cosmic acceleration has been traced mainly  
by SNIa, hence the Hubble relation (distance modulus versus $z$) 
covers the following redshift range $0< z< 1.5$ \cite{Suzuki:2011hu,Betoule:2014frx}.
Also, the binned Pantheon SNIa dataset of 
\cite{Scolnic:2017caz} lies in the range $0<z< 1.6$. 
Moreover, the geometrical probes utilized to study the cosmic
expansion history involve a combination of standard candles 
(SNIa), standard rulers [clusters,
CMB sound horizon detected through Baryon Acoustic Oscillations
(BAOs; \cite{Blake:2011en,Alam:2016hwk}) and the CMB 
angular power spectrum  \cite{Planck16}]. These observational 
probes trace 
the integral of the Hubble expansion rate $H(z)$, hence they 
provide indirect information of the Hubble expansion 
either up to
redshifts of order $z\simeq 1-1.6$ (SNIa, BAO, clusters) or up to the
redshift of recombination ($z\sim 1100$). Clearly, 
the redshift interval $\sim 1.6-1000$ is not
directly probed by the above cosmological data, 
and as shown in \cite{Plionis:2011jj} the redshift range $1.6<z<3.5$
plays an important role in cosmic expansion, since different 
cosmological models give 
their largest differences in this interval
as far as the equation-of-state parameter is concerned.
Owing to the fact that 
the direct $H(z)$ measurements 
can be computed relatively easily at high redshifts
make them, especially those which are located   
at redshifts $z>1.6$, useful tools in 
these kind of studies.
It is interesting to mention that 
there are proposed techniques 
which could expand the direct 
direct measurements of the 
Hubble expansion to $z\le 5$
\cite{Corasaniti:2007bg}.

To date, a demerit of utilizing alone the present  
$H(z)$ data-set in constraining the cosmological models, including 
those of modified gravity, 
is related with the weak statistical constraints due to 
small number statistics. 
However, in order to understand the impact of the 
current $H(z)$ sample in constraining the models, we 
have shown that 
our combined $H(z)$/SNIa/QSO/CMB$_{\text {shift}}$ 
statistical analysis (not affected by $H_{0})$) 
correctly reveals the cosmic expansion 
as provided by the team of Planck
\cite{Planck16}.
Indeed, in the case of $\Lambda$CDM model we found 
$\Omega_{m0}=0.305 \pm 0.001$ which is in excellent agreement with that  
of Planck 2016 TT+lowP+lensing data, namely 
$\Omega_{m0}=0.308\pm 0.012$.

Finally, 
using large sets of Monte Carlo realizations
we studied, for the first time, the ability of
future measurements of the Hubble expansion
to test the viability of the $f(T)$ gravity models.
Interestingly, the outcome of the Monte Carlo analysis
suggests substantial improvement of
the parameter space when we have $\sim 100-120$ $H(z)$ future measurements.
From the observational point of view this is a very realistic prediction
in the context of next generation of surveys.

%%%%%%%%%%%%%%%%%%%%%%%%% Acknowledgments %%%%%%%%%%%%%%%%%%%%%%%%%%%%%%%%%%%%%%%%%%%%
\acknowledgments
 S. Basilakos acknowledges support by the Research Center for Astronomy of the Academy of
Athens in the context of the program ''Testing general relativity on cosmological scales''
(ref. number 200/872). S.~Nesseris acknowledges support from the Research Project
FPA2015-68048-03-3P [MINECO-FEDER], the Centro de Excelencia Severo Ochoa Program
SEV-2016-0597 and from the Ram\'{o}n y Cajal program through Grant No. RYC-2014-15843.
E.N. Saridakis and F. Anagnostopoulos acknowledge support by COST Action (European
Cooperation in Science and Technology) ``Cosmology and Astrophysics Network
for Theoretical Advances and Training Actions''.

%%%%%%%%%%%%%%%%%%%%%%%%%%%%% References %%%%%%%%%%%%%%%%%%%%%%%%%%%%%%%%%%%%%%%%%

\providecommand{\href}[2]{#2}\begingroup\raggedright\endgroup
%%%%%%%%%%%%%%%%%%%%%%%%
\end{document}